# Near-field infrared nano-spectroscopy of surface phonon-polariton resonances

P. McArdle[1,*], D. J. Lahneman[1,*], Amlan Biswas[2], F. Keilmann[3], and M. M. Qazilbash[1,†]

[1]Department of Physics, College of William & Mary, Williamsburg, VA 23187-8795, USA

[2]Department of Physics, University of Florida, Gainesville, FL 32611, USA

[3]Fakultät für Physik & Center for NanoScience (CeNS), Ludwig-Maximilians-Universität, Geschwister-Scholl-Platz 1, 80539 Munich, Germany

**We present combined experimental and numerical work on light-matter interactions at nanometer length scales. We report novel numerical simulations of near-field infrared nanospectroscopy that consider, for the first time, detailed tip geometry and have no free parameters. Our results match published spectral shapes of amplitude and phase measurements even for strongly resonant surface phonon-polariton (SPhP) modes. They also verify published absolute scattering amplitudes for the first time. A novel, ultrabroadband light source enables near-field amplitude and phase acquisition into the far-infrared spectral range. This allowed us to discover a strong SPhP resonance in the polar dielectric $SrTiO_3$ (STO) at an ~ 24 $\mu$m wavelength of incident light.**

## I. INTRODUCTION

Strongly confined electromagnetic fields can couple to collective excitations in materials, such as phonons and plasmons to form polaritons. Polariton wavelengths can be much shorter than the polariton-exciting photon wavelengths. In this way, polaritons are able to confine light to subwavelength regions [1-2]. The ability to control and manipulate electromagnetic energy at the nanometer length scale has the potential for many applications such as photonic computation, nanoimaging devices, and electronic miniaturization [3-6]. In metals and semimetals, coupling to free charge carriers produces plasmon-polaritons. For metals the coupling occurs generally in the near-infrared or visible spectral regions whereas the semimetal graphene has been shown to exhibit a plasmon resonance that can tune through the mid- and far-infrared [1-2,7]. The challenge facing plasmonic applications is the high energy loss which limits plasmon-polariton lifetimes and weakens their resonance [8]. In contrast, a low-loss polariton can form in the infrared reststrahlen band ($\varepsilon_1 < 0$) of polar dielectrics. In this spectral window, polar dielectrics exhibit high reflectance and low loss which leads to the coupling of electromagnetic fields to phonons at the surface, to form surface phonon-polaritons (SPhPs). SPhPs provide an attractive alternative to high-loss plasmon-polaritons. In the mid-infrared ($\approx$ 500 - 4000 cm$^{-1}$), materials such as silicon carbide and boron nitride exhibit SPhP resonances that have been studied with near-field methods [9-12]. However, up to this point, SPhPs in the farinfrared ($\lesssim$ 500 cm$^{-1}$) have been less explored with near-field methods due to the lack of readily available light sources and detectors [8].

---

[*] These authors contributed equally towards this work

[†] mmqazilbash@wm.edu

Traditional infrared spectroscopy is constrained by the Abbe diffraction limit to a minimum attainable spatial resolution of approximately half a wavelength ($\lambda/2$). In the infrared regime, spatial resolution is thus severely limited by the probing wavelength. Near-field infrared microscopy and spectroscopy allow for circumvention of the diffraction limit and provide a nondestructive method of nanometer-scale spatial resolution across the entire visible and infrared spectrum [13-15]. Scattering-type scanning near-field optical microscopy (s-SNOM) employs radiation scattered by the scanning probe tip of an atomic force microscope (AFM). Strong near fields are induced at the tip apex which interact with a sample underneath. The tip scatters radiation following this near-field interaction with the sample, and a detector measures the scattered radiation in the far-field. The AFM tip is used in tapping mode to separate the near-field interaction from background contributions [15]. With this technique, nanometer scale optical properties can be studied with spatial resolution that is only limited by the radius of curvature of the tip apex [16-22]. As a result of the high field confinement, the tip can provide the necessary momentum to resonantly excite SPhPs in dielectrics [22].

Extracting useful information from the near-field infrared experimental data on materials with strong SPhP resonances requires numerical modeling because in this case the near-field infrared interaction between the tip and sample is too complicated to be solved with a closed-form solution. This is due in part to the vastly different length scales in the problem and the role of the probe shaft on the enhancement of the near-field signal. In the past, simplifying approximations have been made in order to make the problem tractable. These involve substituting the probe geometry for an approximate equivalent (either a sphere or an ellipsoid). Here we demonstrate a novel numerical technique to accurately simulate near-field amplitude and phase contrast for the probe-sample system that accounts for detailed probe geometry. Our numerical method is indispensable when a strong coupling between probe and sample exists [23]. This coupling can affect both the spectral position and amplitude of the observed electromagnetic resonances. We have verified the accuracy of our numerical technique versus experimental data on the SPhP resonances present in amorphous $SiO_2$, $Si_xN_y$, and single crystal $SrTiO_3$ (STO). In STO, we discover a strong SPhP resonant mode at an incident wavelength ~ 24 $\mu$m in addition to a weaker one at ~ 15 $\mu$m that has been reported in previous works [24-27]. We were able to observe the strong SPhP resonant mode at low infrared frequency with the use of the ultrabroadband, table-top argon plasma light source developed in-house and by employing a wide-band mercury cadmium telluride (MCT) photoconductive detector. The plasma light source and MCT combination along with a KRS-5 beamsplitter allows us to explore near-field spectra down to 400 cm$^{-1}$ frequency (25 $\mu$m wavelength).

## II. EXPERIMENTAL SETUP

The infrared nanospectroscopy setup consists of our home-built argon plasma light source and a commercial s-SNOM instrument from Neaspec GmbH [24]. The argon plasma light source is an electrode stabilized plasma that is housed inside a sealed water-cooled aluminum vessel with an infrared transparent window to allow access to mid- and far-infrared light emitted from the plasma via bremsstrahlung radiation. We use a potassium bromide (KBr) window for acquiring the Au and STO data and a ZnSe window for the SiO$_2$ and Si data [24]. The vessel is pressurized with 3-15 psi (gauge) of high purity argon gas. A high voltage pulse between the two electrodes ignites an arc discharge that is sustained with a current of about 7 A.

Unpolarized light from the hot spot of the argon plasma is collected and collimated by an off-axis parabolic (OAP) mirror with a 2-in focal length. It is then reflected at a 45° angle of incidence off an indium tin oxide (ITO) coated glass mirror. This transmits much of the unwanted near-infrared and visible radiation and reflects the mid- and far-infrared radiation from the plasma. The reflected beam is then focused through a 500 $\mu$m pinhole by an OAP mirror with a 4-in focal length. This helps to improve the spatial coherence of our beam and ensures we are collecting light from the hot spot of the plasma. The beam is then recollimated using an OAP mirror with a 1-in focal length setting the beam diameter to about 10 mm. We measure a beam power of 1.3 mW after the pinhole in the spectral range between 400 and 5800 cm$^{-1}$. The s-SNOM setup is based on an atomic force microscope (AFM) employing a metal-coated AFM tip and an asymmetric Fourier transform infrared (FTIR) interferometer. The incoming collimated beam is incident on a KRS-5 or ZnSe beamsplitter for either a 400 cm$^{-1}$ lower cutoff or 500 cm$^{-1}$ lower cutoff respectively. The beamsplitter reflects part of the beam to a movable reference mirror and transmits the other part to a parabolic mirror that focuses the beam onto a platinum-iridium coated AFM tip with a radius of curvature of either ~ 20 nm for the SiO$_2$/Si spectra or ~ 60 nm for the STO/Au spectra. The scattered signal is then collected with the same parabolic mirror and is recombined with the reference beam at the beamsplitter and brought to a focus at the detector. For the Au and STO data we use an Infrared Associates (FTIR-22-0.100) MCT photoconductive detector with an active area of 1x10$^{-4}$ cm$^2$, a noise equivalent power of 0.84 pW-Hz$^{-1/2}$, a spectral bandwidth of 400 – 5000 cm$^{-1}$, and a preamp with a 1 MHz bandwidth. This liquid-nitrogen-cooled MCT photoconductor element is housed in a Dewar with a KBr window with moisture-resistant coating. For the SiO$_2$/Si sample, we use a liquid nitrogen cooled photovoltaic MCT detector (Kolmar KLD-0.1- J1/208) with a ZnSe window and a spectral bandwidth of ~ 800 – 3000 cm$^{-1}$. The AFM is operated in tapping mode with a tip oscillation frequency $\tilde{v} \approx$ 250 kHz. A typical tapping amplitude is 80-90 nm for the 60 nm radius tip, and 70 nm for the 20 nm radius tip. The tip oscillation modulates the scattered infrared signal and allows suppression of the background when demodulated at harmonics n$\tilde{v}$ of the tip oscillation frequency $\tilde{v}$, where the demodulation order n = 2, 3, 4 [28,29]. After demodulation, we obtain a scattering amplitude s$_n$ and scattering phase $\phi_n$. To eliminate unwanted signal fluctuations caused by atmospheric absorption lines we encased the entire beam path and microscope in a dry and CO$_2$-free air purge.

### III. SIMULATION METHOD

Modeling of the near-field interaction has previously been accomplished with a simplification of the probe geometry in which the probe is replaced with a sphere at the tip's apex [28,30]. In the electrostatic approximation $a \ll \lambda_{\text{inc}}$, where $a$ is the radius of the sphere of the order of 10 nm, the effective polarizability over an infinite substrate as a function of gap distance has an analytic form. The scattered field $E_{\text{scat}}$ is proportional to the effective polarizability $\alpha_{eff}$

$$E_{scat} \propto \alpha_{eff} = \frac{\alpha(1+r_p)^2}{1-\frac{\alpha\beta}{16\pi(a+z)^3}} \quad (1)$$

where $\alpha = 4\pi a^3 \frac{\varepsilon_t - 1}{\varepsilon_t + 2}$, $\beta = \frac{\varepsilon - 1}{\varepsilon + 1}$, $r_p$ is the Fresnel reflection coefficient of the sample for p-polarized light, and z is the gap distance between the probe apex and the sample surface. Note that $\alpha$ is the polarizability of an isolated sphere with complex dielectric function $\varepsilon_t$, and β is the response function of the material with complex dielectric function ε. Since the near-field component of the scattered field is nonlinear in the gap distance z, one can separate the near-field contribution to the scattered radiation from the background contributions by modulating the gap distance [28,30]. Demodulation at multiples of the probe modulation frequency provides the near-field amplitude and phase. A resonance condition can come from the sample and the sphere. When $\varepsilon_1$ = -1, it forces the response function β to a maximum and generates a sample resonance. Candidate materials that have regions of negative $\varepsilon_1$ are metals/semimetals near their plasma resonance or polar dielectrics near the optical phonons. Polar dielectrics exhibit metal-like properties in regions called reststrahlen bands in the infrared [1-2]. The reststrahlen bands exist between longitudinal and transverse optical phonons. These regions have high reflectance and low loss which can lead to the coupling of confined electromagnetic fields and phonons to generate SPhPs. To quantify SPhP resonances, the Q factor defined as full-width at half-maximum divided by the center frequency of the resonance peak is often used as a measure of quality [31]. An SPhP mode with large Q represents a low loss, narrow bandwidth mode whereas the opposite is true at low Q. For surface plasmons in metals, the highest Q reported is around 40 [32]. In contrast, higher

Q has been experimentally demonstrated for SPhP resonances [31,33-34], and can be expected about an order of magnitude higher from theory [1].

The probe can strongly couple to optical phonons in the sample and generate SPhP modes. The probe-sample resonance depends both on the local material dielectric function and the probe geometry. The simple description given above that approximated the probe tip as a small sphere (point dipole model) can approximately describe SPhP resonances [22] but is certainly inadequate at reproducing all observed features. To analytically improve the point dipole model one can replace the probe with an ellipsoid [30]. The ellipsoid acts as a finite dipole, with charge densities accumulating at the caps and improves the point dipole model by considering an approximate probe shaft. The finite dipole model more accurately fits experimental approach curves compared to the point dipole model, and therefore the finite dipole model is preferred in the analysis of experimental data [30]. While the finite dipole model is preferred over the point dipole because of its improved accuracy for describing resonant materials, it relies on phenomenological parameters which are not unique: Different sets of parameters can lead to the same fits to an approach curve (see Appendix A). Other quasianalytical models have also been developed, in which a portion of the problem is numerically computed and used as part of an analytic solution that remains subject to underlying assumptions [23,35]. Here we introduce a full-wave numerical simulation that considers the probe in sufficient geometric detail to explain the experimental data quantitatively. Our numerical approach in this work provides an alternative modeling method to the analytical and quasianalytical methods discussed above. Our numerical approach does not rely on adjustable, phenomenological parameters, and can be used to model a variety of near-field infrared data sets in different experimental conditions. Previous numerical studies did not exploit their full capabilities [36-39] as they used truncated shaft lengths in the assumed probe geometries to reduce computational complexity. The shaft length which is ~ 15 µm for commonly available commercial AFM probes becomes a critical parameter in the quantitative simulation of mid- and far-infrared s-SNOM of SPhP resonances because the shaft length is similar to the incident wavelengths. Using the correct length and shape of the shaft was essential to obtain accurate numerical results in our spectral range of interest (400 - 1200 cm$^{-1}$).

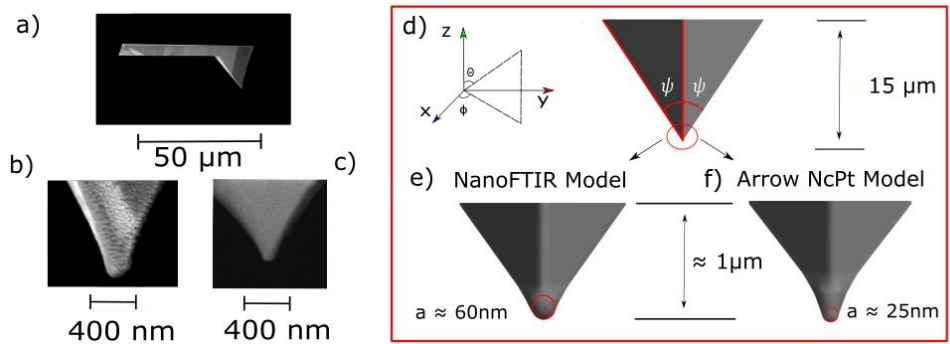

**FIG 1**. (a) SEM image of the entire AFM nano-FTIR probe consisting of tip, shaft, and cantilever. (b) Higher- resolution SEM image of the nano-FTIR tip apex. (c) High-resolution SEM image of the Arrow tip whose apex has a smaller radius of curvature compared to the nano-FTIR tip. d) Tetrahedral shaft with full angle $\psi \sim 60°$ based on the SEM images, which is used in both probe models. The tip apex within the red circle is shown in panels (e), (f). (e) Geometry of the nano-FTIR tip apex with radius of curvature a ~ 60 nm. (f) Geometry of the Arrow tip apex with radius of curvature a ~ 25 nm.

The simulations were performed with FEKO, a proprietary computational electromagnetic solver for arbitrary bodies. It has an embedded computer aided design (CAD) modeling interface and allows for the importing of externally generated models [40]. With this capability we generated an accurate model of our AFM probes. Our experimental probes are ~15 µm tall tetrahedral shafts from tip apex to the cantilever arm and have an approximate front and side full angle of $\psi \sim 60°$. To properly include the curvature at the tip apex, we obtained scanning electron microscopy (SEM) images of the probe and extracted a two-dimensional (2-D) outline of the apex from our SEM images. It extends approximately 150 nm up the tip length. A circular sweep of the outline was performed to generate a three-dimensional point cloud. This was fit to a nonuniform rational B-spline (NURBS) to form a closed surface. This was then connected smoothly to our tetrahedron with additional NURBS surfaces. Altogether, the geometric model is a detailed replica of the shaft and tip of our probe. The cantilever has been omitted from the geometric model because it has limited influence on the simulated tip-sample near-field interaction. We generated two models both with the same tetrahedral shaft, but with different tip apex radii. We took SEM images for each probe modeled: one is a Neaspec nanoFTIR probe with tip apex radius ~60 nm and the other is an Arrow NCPt probe with tip apex radius ~25 nm. From our SEM images we found that the same tetrahedron could be used for both models which are displayed in Fig 1. The commercially available AFM tips generally used in experiments are composed of silicon coated with a thin metallic layer. To reduce computational complexity in our numerical simulations, we assumed the tip is a perfect electrical conductor (PEC). We demonstrate later in the paper that using a PEC tip gives nearly identical results when compared to simulations with a metal-coated silicon tip.

The simulation methods used were the method of moments (MoM) and surface equivalence principle (SEP) coupled with a half-planar/multilayered Green's function. The MoM is well suited for solving radiation and scattering problems [41]. Only the scatterer's surface is meshed which greatly reduces the computation time as compared to finite difference time domain (FDTD) or finite element method (FEM) solvers which have volumetric meshes. The total induced current and charge is represented by these surface mesh sites which determine the scattered field [42-44].

As compared to the FDTD method, the MoM employs a curvilinear mesh which allows for minimal but accurate meshing of curved surfaces. Application of the MoM also allows the use of a planar or multilayered Green's function which is an accurate and efficient approach to computing near-field contrast of bulk isotropic materials. Using the FDTD or FEM methods would require simulating a finite sized substrate and absorbing boundary conditions which greatly increases the computational complexity [44]. In the limit of an infinitesimal mesh size, the MoM provides an exact solution to the full wave equation. The mesh size can be chosen to be both accurate and efficient.

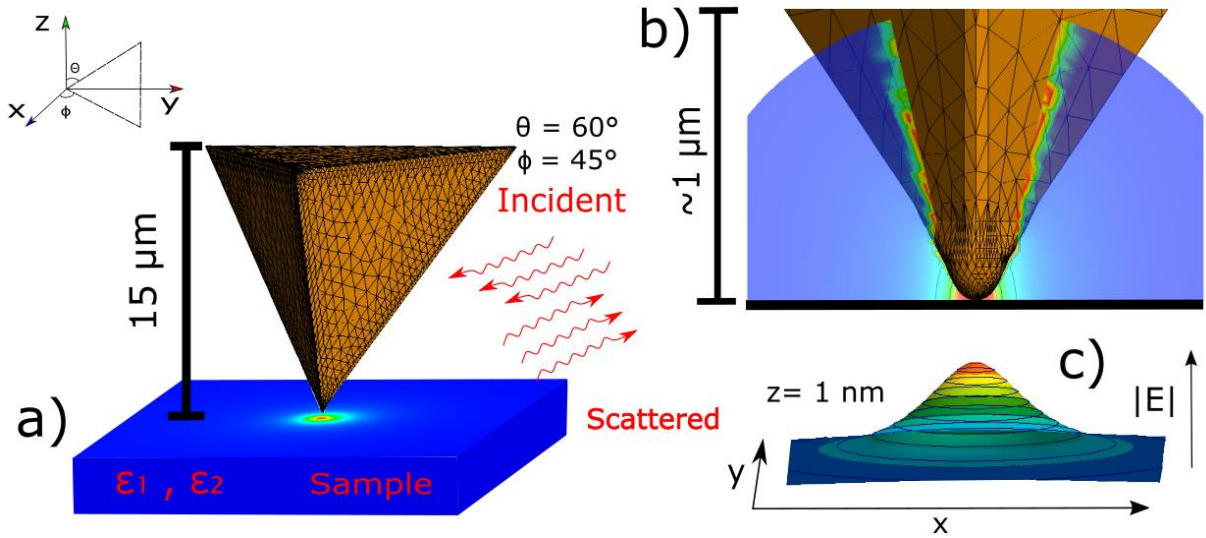

**FIG 2**. (a) A schematic representation of the simulation showing the meshed AFM nano-FTIR probe, the sample, and the incident and scattered electromagnetic fields. The sample's dielectric function has real and imaginary parts $\varepsilon_1$ and $\varepsilon_2$, respectively. A fine mesh was used near the tip apex and a coarse one far from it. (b) Front side schematic of the AFM probe near the tip apex over an infinite Au sample and the computed electric field distribution. The probe shown in panels (a,b) is meshed as it was in simulations presented in this work. (c) Plot of the computed electric field distribution as a function of x-y spatial position at a fixed z position of 1 nm over an infinite Au sample with the probe at a height of 10 nm above the sample. The electric field is computed at a frequency of ~438 cm$^{-1}$.

In our simulations, we employed the probe geometries described earlier in this paper above a half-planar sample. A plane wave is incident at $\phi = 45°$, $\theta = 60°$ and extends throughout the solution space which is a 1 mm x 1 mm x 1 mm cube. The simulation setup is illustrated in Fig 2. Experiments generally employ a convergent incident beam with an angular spread of about $\pm 20°$. Instead of a convergent beam, plane waves were used in the majority of the simulations in this work to reduce computational complexity. We demonstrate that simulations using this simplification provide a good quantitative description of the experimental spectra when the experimental spectra are normalized to a reference material such as gold (Au) or silicon (Si), or when the experimental spectra are obtained without normalizing to a reference material. However, for simulations that quantitatively describe the absolute scattering amplitude measurements of Amarie and Keilmann [45], it is important to include details of the incident beam's characteristics and we have implemented a convergent beam with a diffraction limited spot size.

To give an example, we have simulated the electric field distribution underneath our model's tip apex above an infinite Au substrate at a fixed height and wavelength as a function of lateral position, and these results are presented in Fig 2. The incident plane wave has *p*-polarization in the numerical simulations presented in this work, such that most of the electric field is along the long axis of the probe shaft. It is well established that the near-field enhancement comes primarily from the electric field along the long axis of the probe shaft [46, 47]. Our numerical simulations with *p*-polarized light and *s*-polarized light confirm that the demodulated s-SNOM signal is dominated by contributions from *p*-polarized light. The incident plane wave illuminates the tip-sample system and the scattered far-field is computed in the same direction as the incident plane wave. The scattered fields are simulated across a specified spectral bandwidth and with the tip at different discrete positions above the sample. These positions are determined by taking into account the tapping amplitude and frequency used in the experiments. Having simulated the scattered field as a function of tip position and frequency, the following parametrization is used to mirror the fields in time in order to make them periodic.

$$z(t) = \frac{A}{2}(1 + cos(\Omega t)) \quad (2)$$

Here $\Omega = 2\pi\tilde{\nu}$ and $\tilde{\nu}$ is the tip oscillation frequency. The incident and scattered electric fields have frequency dependence $\nu$ and the scattered electric field is periodic in time due to the oscillating tip. Hence, one can expand the scattered electric field into a Fourier series

$$E_{scat}(t, \nu) = \sum_{n=0}^{\infty} c_n(\nu) * e^{i2\pi n \tilde{\nu} t} \quad (3)$$

Employing the Fourier transform method, one can solve for the complex valued coefficients

$$c_n(\nu) = \frac{1}{T}\int_0^T E_{scat}(t, \nu) e^{-i2\pi n \tilde{\nu} t} dt \quad (4)$$

Where $T = 2\pi/\Omega$ is the period of tip oscillation. From these coefficients one obtains the field amplitude $s_n$ and phase $\phi_n$ that are sampled in the far-field

$$c_n = s_n e^{i\phi_n}, \quad s_n = |c_n|, \quad \phi_n = tan^{-1}\left[\frac{Im(c_n)}{Re(c_n)}\right] \quad (5)$$

For referenced near-field amplitude and phase contrast, the simulation is performed on both a sample of interest and a reference material. This results in the following near-field contrasts

$$\frac{c_n^{sample}}{c_n^{ref}} = \frac{s_n^{sample}}{s_n^{ref}} e^{i(\phi_n^{sample} - \phi_n^{ref})} \quad (6)$$

Where the ratio $\frac{s_n^{sample}}{s_n^{ref}}$ is the demodulated near-field amplitude contrast and $(\phi_n^{sample} - \phi_n^{ref})$ is the demodulated near-field phase contrast.

Computations were done primarily on the high performance computing (HPC) center at the College of William & Mary. Depending on the complexity and size of a particular simulation, between one and four nodes of a subcluster containing between eight and 16 cores/node were used. The complexity depends on the number of tip positions simulated and spectral bandwidth. Generally, between six and 20 tip positions were simulated and the bandwidths on the order of hundreds of cm$^{-1}$. For example, for simulations on STO and Au, later presented in the Results section, we simulated data from 300 to– 1000 cm$^{-1}$ at 12.5 cm$^{-1}$ spectral resolution, which equates to 56 frequencies. Twelve tip positions were simulated for each frequency and two samples (STO and Au) were modeled, for a total of 1344 simulations. On a desktop computer (Dell Optiplex 9020, processor: Intel i7-4790 CPU @ 3.6 GHz (four cores), installed memory: 16 GB RAM), a single frequency and tip position for the model in this work can be completed in ~ 3 min. If one were to do the STO and Au simulations one at a time on a desktop computer, it would take ~ 67 h or just under 3 days. However, with the HPC we are able to parallelize the computations, which provides the most dramatic reduction of a factor of about 10 in the simulation times compared to a desktop computer.

## IV. RESULTS AND DISCUSSION

Since we simulated the detailed probe geometry, the computed scattered far- and near-field signals exhibit antenna resonances [48]. These resonances correspond to charge oscillations of either even or odd charge symmetry at the antenna's end [49], and are plasmonic in origin, coming from the metallic layer of our AFM tip. The scattering from the AFM probe also has a strong angular dependence which is influenced by these antenna modes consistent with a previous numerical study [50]. These aspects are further discussed in Appendixes B and C.

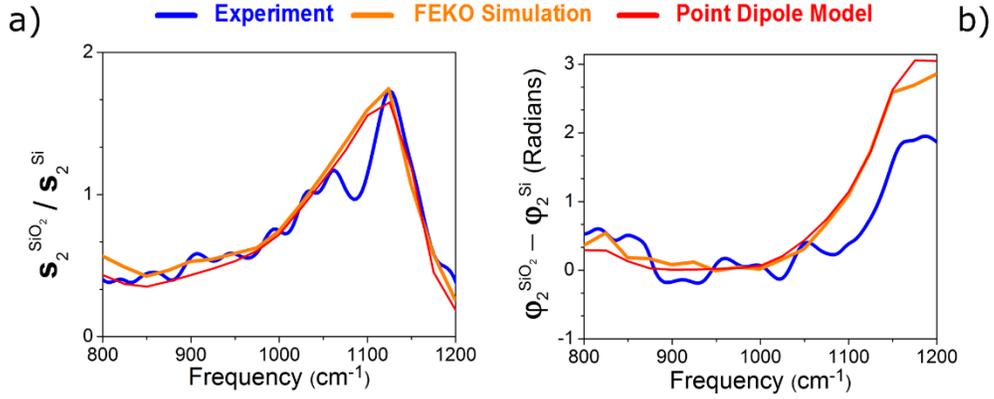

**FIG 3**. (a) The n = 2 near-field amplitude spectrum and (b) phase spectrum of 100-nm-thick $SiO_2$ on Si normalized to the spectra of Si. The experimental data are shown along with results from the point dipole model and numerical FEKO simulations.

With the spectral response of the probe model characterized, we first use our method to simulate the well-studied phonon-polariton resonance in amorphous $SiO_2$ film that we experimentally measured in our s-SNOM setup. The sample is composed of a 100 nm layer of thermally grown $SiO_2$ over a bulk Si substrate. The real and imaginary parts of the complex dielectric functions used to simulate $SiO_2$ and Si are plotted in Appendix E. The tapping amplitude was 70 nm and the tapping frequency was 250 kHz, both fixed by experiment [24]. The Arrow NCPt tip model was used for the simulation consistent with the AFM probe used in the experiment. As seen in Fig 3, both the simulated data and point dipole model match the experimental data well. This can be attributed to the relatively weak SPhP in amorphous $SiO_2$ due to high damping, such that the calculated $SiO_2$ spectra when referenced to the calculated Si spectra are nearly independent of the probe geometry [22].

To further benchmark our simulation method, simulations were performed to model s-SNOM data on SPhPs in materials with isotropic dielectric function. We used previously published s-

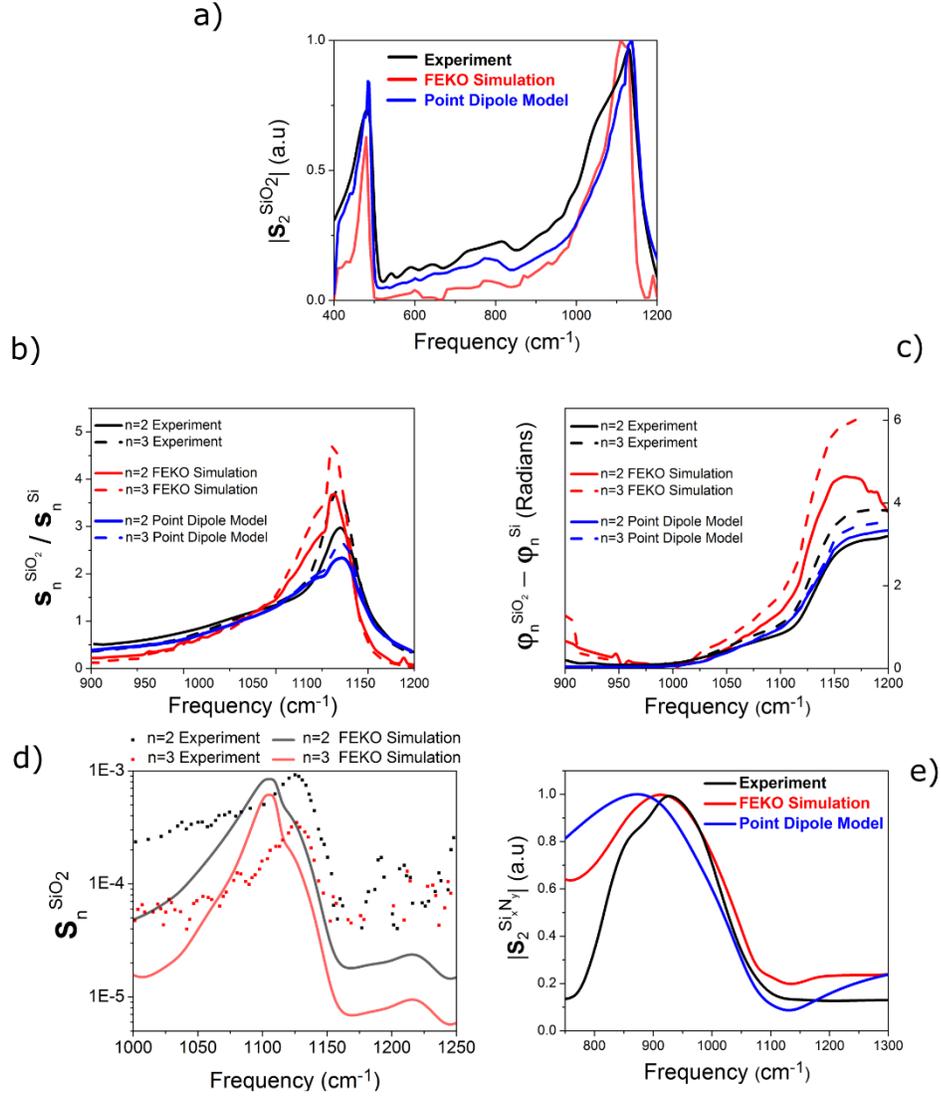

**FIG. 4**. Experimental and simulated infrared near-field spectra on amorphous $SiO_2$ and amorphous $Si_xN_y$. (a) The n = 2 near-field amplitude for bulk, amorphous $SiO_2$. The experimental data are taken from Ref. [51]: (b-c) The n = 2, 3 near-field amplitude and phase of 300-nm amorphous $SiO_2$ on Si referenced to Si. The experimental data are taken from Ref. [23]. (d) Absolute n = 2, 3 near-field amplitude of bulk, amorphous $SiO_2$. The experimental data are taken from Ref. [45]. (e) The n = 2 near-field amplitude of 40-nm $Si_xN_y$ on Si substrate. The experimental data are taken from Ref. [52].

SNOM data obtained on amorphous $SiO_2$ and amorphous $Si_xN_y$ by other groups, and s-SNOM data measured on single-crystal STO in our experimental setup. The dielectric functions of these materials and the gold reference are plotted in Appendix E. The simulations took into account the experimental tapping amplitude and frequency of the AFM tip. The numerical simulations are plotted in Fig 4 along with the point dipole model calculations and the experimental data.

A broadband spectrum of bulk amorphous $SiO_2$ which extends down to ~ 400 $cm^{-1}$ [51] is interesting because $SiO_2$ supports an additional SPhP mode at ~ 450 $cm^{-1}$. The n = 2 amplitude measured in the experiment is plotted in Fig. 4(a) along with the simulated spectrum. Note that the experimental data in Fig. 4(a) were not normalized to a reference material. Nevertheless, a reasonably good match is seen between the simulation and experiment. We next simulated the n=2 and n=3 amplitude and phase for s-SNOM data on another amorphous $SiO_2$ sample [23]. The sample consisted of 300-nm $SiO_2$ on Si and the experimental data on $SiO_2$ were referenced to Si. The NCPt tip model was used in the simulation which differed from the tip used in the experiment [23]. Nevertheless, the FEKO simulation provides a reasonable description of the experimental data [Figs.- 4 (b) and 4(c)], although it slightly overestimates the amplitude of the SPhP. The point dipole model also provides a reasonable match to the experimental data in Figs. 4(a)-4(c) although it underestimates the amplitude of the SPhP resonance in $SiO_2$ when referenced to Si [Figs. 4(b)-4(c)].

Next, we simulated *absolute* scattering efficiencies of bulk, amorphous $SiO_2$ for the first time and compared to them experimental absolute scattering efficiencies previously published by Amarie and Keilmann in Ref. [45]. The absolute scattering efficiency is defined as

$$\sigma_{abs} = \frac{E_{scat}}{E_{inc}} \quad (7)$$

where $E_{scat}$ is the scattered field amplitude from the tip and $E_{inc}$ is the incident field amplitude. By modulating the tip, $\sigma_{abs}$ is decomposed into Fourier components as shown in equations 3-5. We employed a half-planar Green's function as the SiO$_2$ sample and the Arrow NCPt tip model for the simulation. For computing the absolute scattering amplitude, it is essential to use the detailed probe geometry as well as the incident beam characteristics consistent with experimental conditions [45]. We used a convergent, incident beam with a focusing half angle of 26°. The spot at the focus is diffraction limited and has a size ~10 μm (beam waist) in the spectral range 1000 -1250 cm$^{-1}$. The backscattered field was calculated over the collection half-angle of 26° and demodulated using Eq. (3)-(5). The demodulated scattered field from the probe-sample system was then divided by the incident field for each wavelength following the normalization procedure used in the experiments. The experimental and simulated results are plotted in Fig 4(d). It is encouraging to see that the simulated absolute scattering amplitude for both n = 2 and n = 3 has the same overall shape compared to the experimental absolute scattering amplitude and is a good match to the experimental data in the spectral range of the SPhP resonance peak. Absolute scattering has been previously calculated using the finite dipole model. The finite dipole model gives a good match to the experimental SPhP frequency but requires the use of a scaling parameter to match the experimental SPhP peak height [45]. The point dipole model gives incorrect results for absolute scattering and is not considered in Fig 4(d).

    To further demonstrate the capabilities of our numerical method for a material other than amorphous SiO$_2$, we simulated a strong SPhP resonance in amorphous Si$_x$N$_y$. We compared our simulation to the experimental data from a 40-nm Si$_x$N$_y$ film reported in Ref. [52]. The NCPt tip model was used in this case. The n=2 amplitude from the simulation is compared with experimental data in Fig .4(e). The experimental data are not referenced to a known sample such

as silicon or gold. Hence, the simulation is in semiquantitative agreement with experiment. In contrast, the point dipole model spectrum is redshifted from experiment. The SPhP resonance of $Si_xN_y$ is less damped than the SPhP modes in $SiO_2$, thus the probe geometry becomes more important for modeling purposes. This can be seen in the dielectric function used for $Si_xN_y$ plotted in Appendix E. Note that the low-frequency detection limit of the experimental data in Fig. 4(e) is ~ 750 cm$^{-1}$. The detector roll-off near 750 cm$^{-1}$ is most likely the cause of the discrepancy at low frequency in the amplitude spectrum between the experiment and the simulations.

For a quantitative benchmark of a strongly resonant sample, we modeled our experimental spectra of STO referenced to gold. Figure 5 shows experimental amplitude spectra of STO taken at the second, third, and fourth harmonics with 12.5 cm$^{-1}$ spectral resolution. As reported previously, there is a broad phonon-polariton resonance centered at ~ 640 cm$^{-1}$ [24-27]. However, our data in this paper extend to lower frequencies compared to previous works. Hence, we discover a large resonant peak occurring at 425 cm$^{-1}$ rising to about 4, 6, and 9 times the amplitude of the gold reference in the second-, third-, and fourth-harmonic signals, respectively. Since the amplitude of the resonance increases between harmonic order n=2 to n=4, it further indicates a near-field resonance arising from surface confinement of the electric field because the higher harmonic orders are more sensitive to the surface [53,54].

We have applied our simulation method to quantitatively describe the experimental results on STO. For this simulation, our Neaspec nano-FTIR tip model was used. The simulation tapping amplitude and frequency were 90 nm and 250 kHz respectively, based on the experimental parameters. Both the Au and STO were simulated with a half-planar Green's function. Simulated near-field amplitude and phase data for STO referenced to Au for n = 2, 3 and 4 are plotted in Fig. 5 together with the experimental data and point dipole model results. For the point dipole model results, the same experimental parameters used in the FEKO simulation were employed. These parameters were a platinum sphere of radius a=60 nm, tapping amplitude A=90 nm, and tapping frequency $\tilde{\nu}$=250 kHz.

Our simulated data agree well with the measured 425-cm$^{-1}$ SPhP mode in absolute height for n = 2. At higher demodulation orders, the peak height of this SPhP mode in the experimental data somewhat exceeds the peak height in the simulation. Nevertheless, our numerical simulations are in much better quantitative agreement with the experimental amplitude data compared to the point dipole model. A dip in the experimental amplitude spectra for n=2, 3 at ~ 610 cm$^{-1}$ is not present in our numerical model. Nor is this dip present in the point dipole model and the finite dipole model. We provide possible reasons for this feature in Appendix D.

We observe that the experimental resonance position occurs at the same $\varepsilon_1$ for both SPhPs (see Appendix E). However, the lower lying SPhP mode (~ 425 cm$^{-1}$) has a factor of 4 lower $\varepsilon_2$, indicating lower damping. The SPhP resonances lead to significant structure in the phase spectra as well (Fig. 5). We see that the simulated phase is in good agreement with the experimental phase for the three demodulation orders. We also see that the point dipole model is inadequate for reproducing the measured phase spectra.

The Q factors for both SPhP modes in STO are consistent between the experimental and simulated spectra. The Q factor values are extracted from the SPhP modes in the n$^{th}$ demodulated near-field amplitude data. The linewidth of the giant mode centered at 425 cm$^{-1}$ frequency decreases with increasing demodulation order and there is negligibly small shift of the center frequency. This phenomenon has been previously reported for midinfrared SPhPs in SiC [8,45]. The 425-cm$^{-1}$ SPhP mode in STO has a Q factor ~8-10 in the demodulated amplitude data. This is lower compared to the Q factor ~ 20 of the midinfrared SPhP in SiC. However, the Q factor of the 425-cm$^{-1}$ SPhP mode in STO is comparable to or higher than the Q factors of the far-infrared SPhP resonances in $Al_2O_3$ and $SiO_2$ [51].

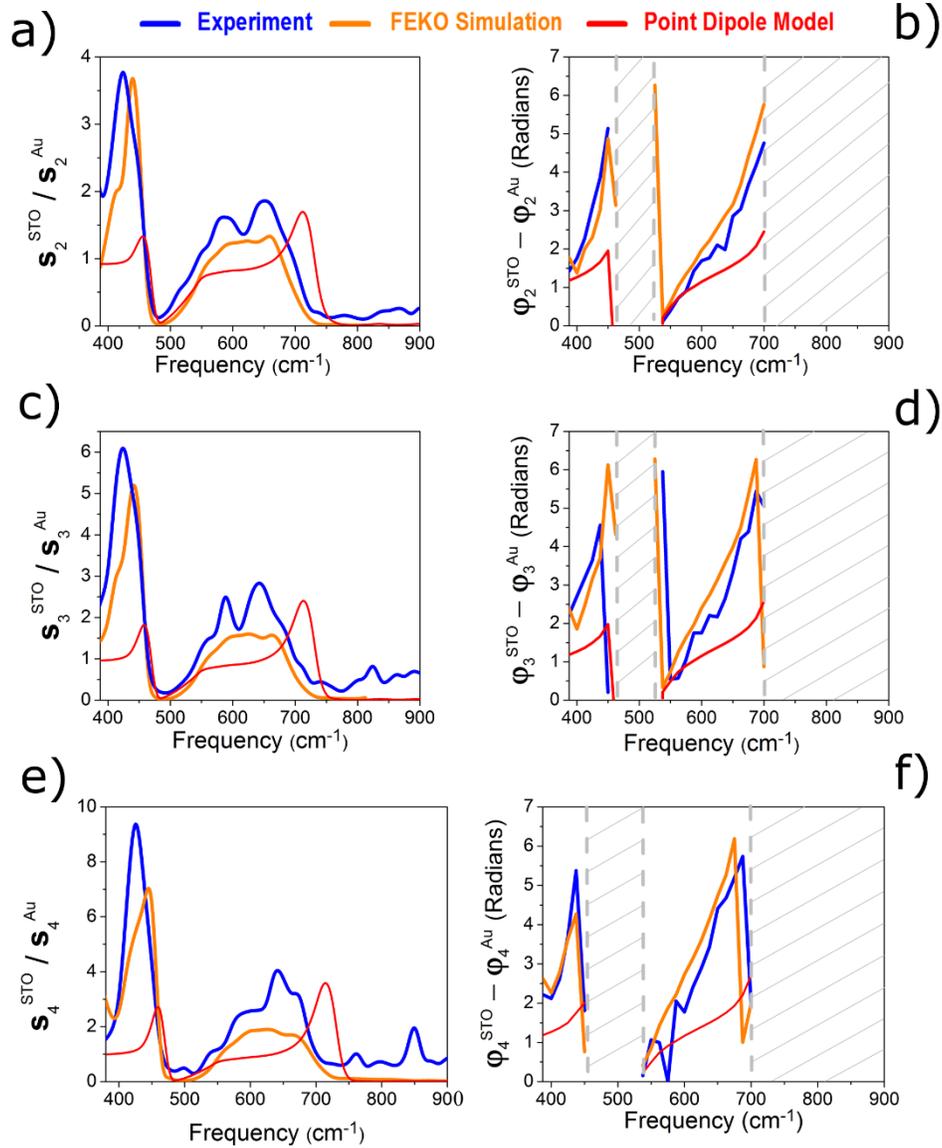

**FIG. 5**. (a-f) Plots show the experimental results, numerical simulations, and point dipole model calculations of the n = 2, 3, 4 near-field infrared amplitude and phase of STO normalized to the spectra on gold. The phase is indeterminate in the spectral regions depicted by the gray hatched areas because the scattering amplitude from STO is negligibly small in these spectral regions.

Tremendous savings in computation can be gained by approximating the tip as a PEC. In the method of moments, the required memory scales as $m^2$ where $m$ is the number of mesh elements. To simulate the probe as a multilayered model with the SEP, the scaling jumps to ~ $4m^2$ [40]. To ensure that the PEC tip simplification does not compromise the accuracy of the numerical simulations, we used a multilayered model of the tip and simulated the near-field contrast of STO and Au. The multilayered tip model consists of a tetrahedral silicon core surrounded by a 50-nm-thick Pt layer. This multilayered model is displayed in Fig. 6. We chose the platinum dielectric function for the simulations instead of the dielectric function of the PtIr alloy because the precise composition of the PtIr alloy is unknown as it is considered proprietary by the manufacturers of the probes used in experiments. Moreover, the literature indicates that the Ir composition present in the probes is significantly less than Pt [55]. The simulation results are not expected to differ as long as the dielectric function of a good electrical conductor is employed for the metallic layer. The dielectric functions used for Si and Pt are plotted in Appendix E. The multilayered tip was simulated using the SEP and the substrate was simulated using a half-planar Green's function. This method was chosen so that direct comparison of the data could be made with the PEC model of the probe over a half-planar Green's function. The near-field amplitude and phase spectra for the multilayered and PEC tips are shown in Fig. 6. The simulated amplitude and phase spectra for the multilayered and PEC tips are consistent to about a few percent. Therefore, we conclude that the PEC tip is an accurate representation of reality for the spectral range studied in this work.

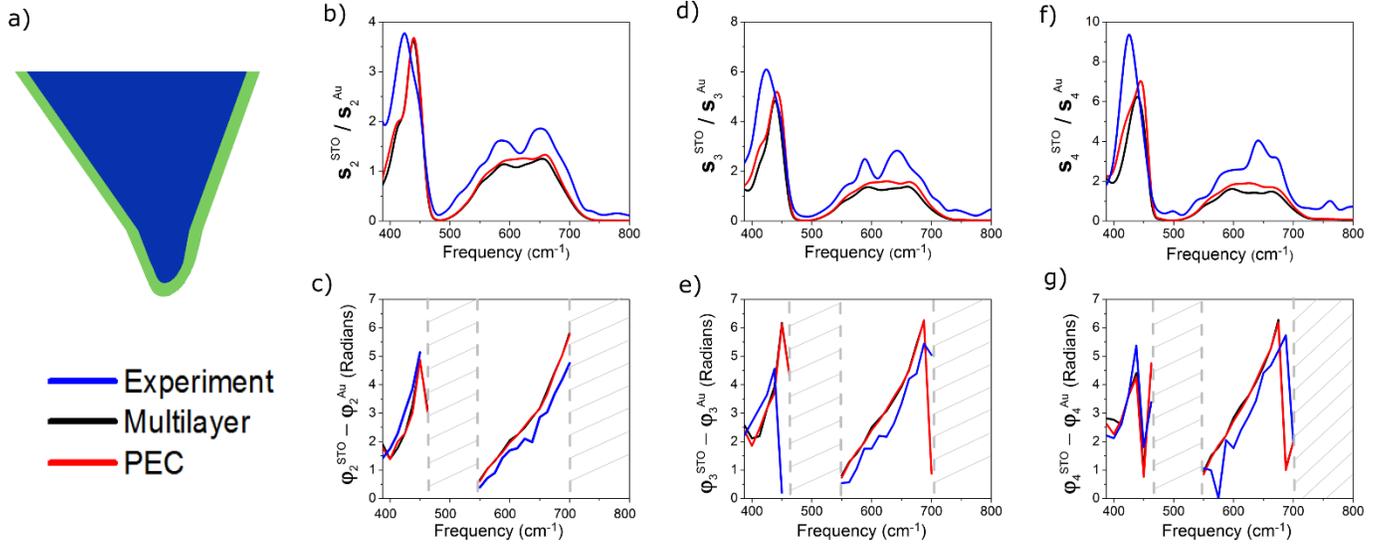

**FIG. 6**. a) Multilayered model of the AFM probe. Blue region indicates silicon core and green region depicts the 50 nm thick Pt coating. b-g) Simulated n = 2, 3, 4 amplitude (first row) and phase (second row) of STO normalized to the spectra on gold for the multilayered tip model and the PEC tip model are compared to experiment. The phase is indeterminate in the spectral regions depicted by the gray hatched areas because the scattering amplitude from STO is negligibly small in these spectral regions.

## V. CONCLUSIONS

In conclusion, we have developed a numerical technique to accurately model the experimental near-field amplitude and phase spectra in near-field infrared nanospectroscopy. This numerical method models the AFM probe geometry in sufficient detail and hence provides a parameter-free, quantitative description of near-field amplitude and phase spectra of surface phonon-polariton resonances. Our numerical method is especially useful for describing surface phonon-polaritons in materials with strong probe–sample coupling. By utilizing a novel, broadband infrared light source, we experimentally observe a strong surface phonon-polariton mode in the polar dielectric

SrTiO$_3$ at ~ 425 cm$^{-1}$ in the far infrared. This resonant mode is quantitatively explained with our numerical method. Extension of our numerical technique to materials with anisotropic dielectric function and/or heterogeneous structure is possible in the future. Numerical simulations are also suggested to elucidate fundamental physical phenomena measured in near-field infrared nanospectroscopy experiments on novel, complex materials.

## ACKNOWLEDGMENTS

M.M.Q. acknowledges support from the National Science Foundation (NSF) via Grants #No. DMR-1255156 and #No. IIP-1827536. A.B. acknowledges support from NSF (Grant #No. DMR-1410237). The simulation work was performed, in part, using computing facilities at the College of William & Mary which are supported by contributions from the National Science Foundation, the Commonwealth of Virginia Equipment Trust Fund, and the Office of Naval Research. M.M.Q. and P.M. acknowledge helpful suggestions from Seth Aubin and Eric Walter concerning the electromagnetic simulations.

P.M and D.J.L contributed equally to this work.

## Appendix A: Finite Dipole Model

The finite dipole model requires a separate discussion which is tangential to the work described in the main text. The finite dipole model requires experimental approach curves to be measured. This is because approach curves must be fit to extract the parameters $g_1$, $g_2$, and L in the finite dipole model [30]. Note that $g = g_1 + ig_2$, where $g_1$ and $g_2$ are the real and imaginary parts of the complex parameter g. There are multiple combinations of $g_1$, $g_2$, and L that adequately fit the

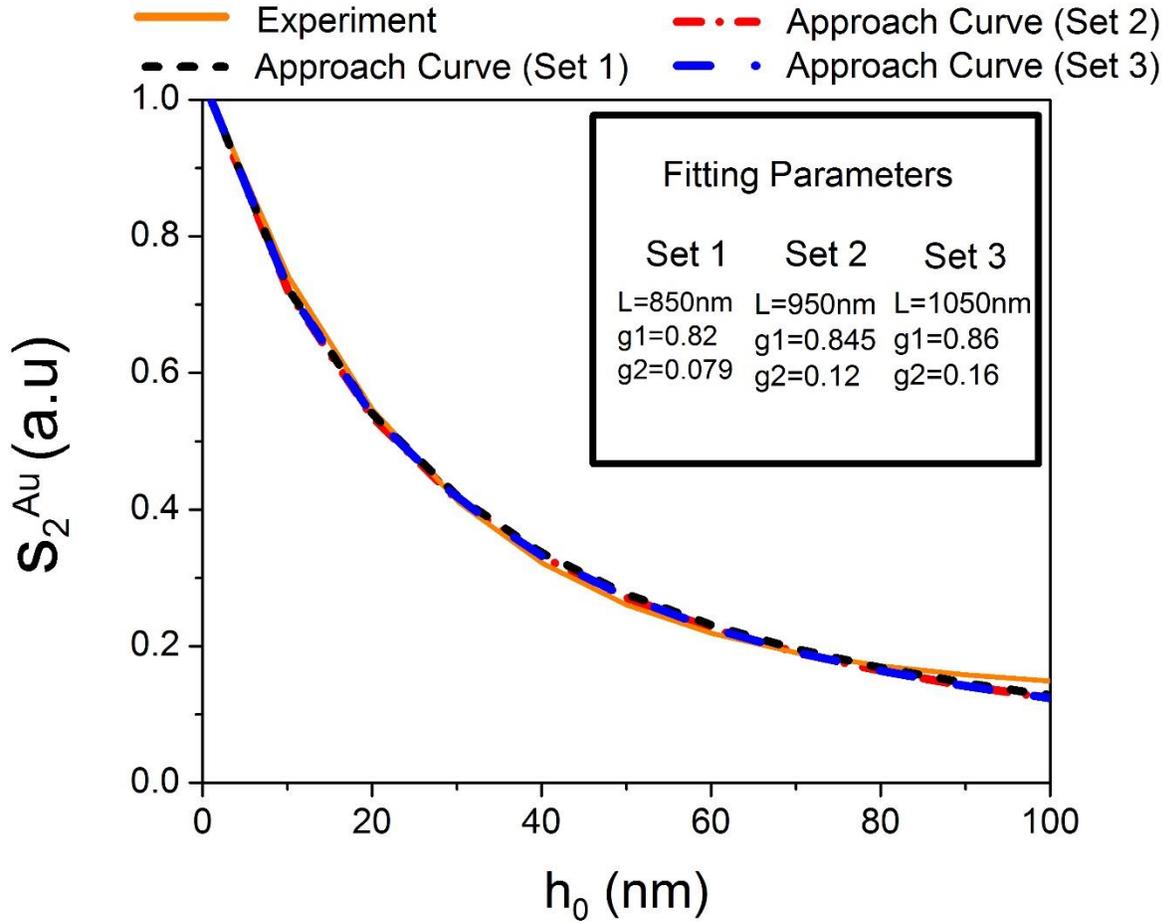

**FIG. 7**. Experimental approach curve obtained with the nano-FTIR probe at the white light position on Au. The n = 2 near-field amplitude is plotted as a function of tip-sample distance ($h_0$). Also shown are fits with three different sets of fitting parameters $g_1$, $g_2$, and L.

approach curve data with differing results in the demodulated near-field amplitude spectra. This can be seen in Fig 7 where, for example, three different sets of $g_1$, $g_2$, and L do a good job of reproducing our experimental approach curve. Using these three different sets of parameters, we calculated near-field amplitude and phase contrasts for STO normalized to gold (Au) which are displayed in Fig. 8. It can be seen in the amplitude spectra that the resonances occur at the same positions but the peak heights are highly dependent on the fitted parameters $g_1$, $g_2$, and L. For the finite dipole model, the calculated spectrum that is closest to the experimental data is displayed along with the experimental data. The finite dipole model requires parameters extracted from approach curves to simulate near-field contrasts and also relies on the measured near-field experimental data for selecting a suitable set of parameters. Our numerical method is not based on extracting and selecting parameters from measured approach curves and measured near-field spectra. For comparison, we have plotted the finite dipole model calculations with parameters from set 1 of Fig. 7 along with our numerical model simulations and experiment in Fig. 9. The set 1 parameters of the finite dipole model yield a reasonably good match to the experimental low frequency SPhP resonance although the width and shape of experimental high-frequency SPhP resonance is not properly captured. A different set of parameters of the finite dipole model can give a good match to the high-frequency SPhP peak height but will underestimate the low-

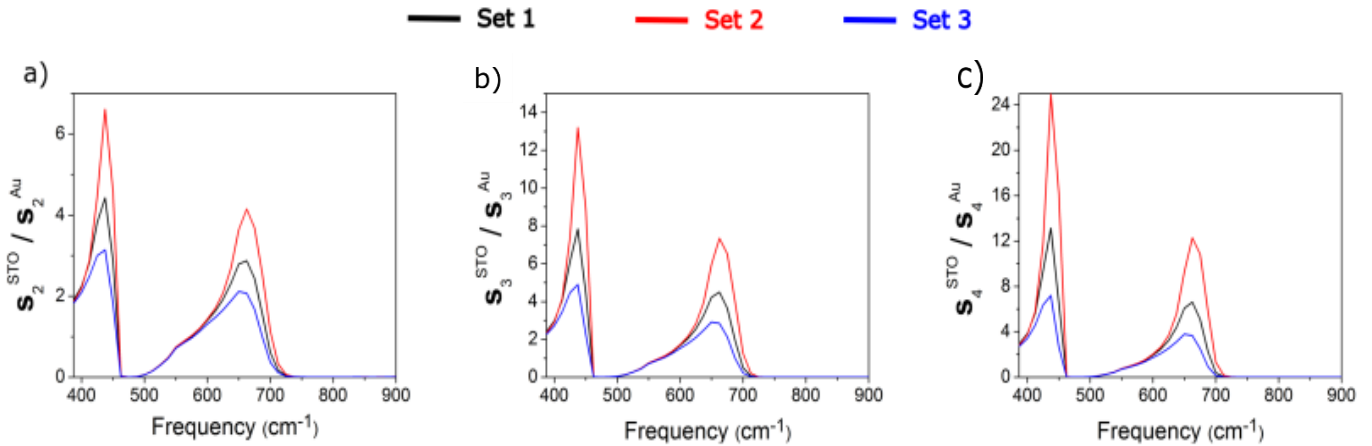

**FIG. 8**. (a-c) The n = 2, 3, 4 near-field amplitude calculated using the finite dipole model for three different sets of parameters $g_1$, $g_2$, and L listed in the inset of Fig. 7.

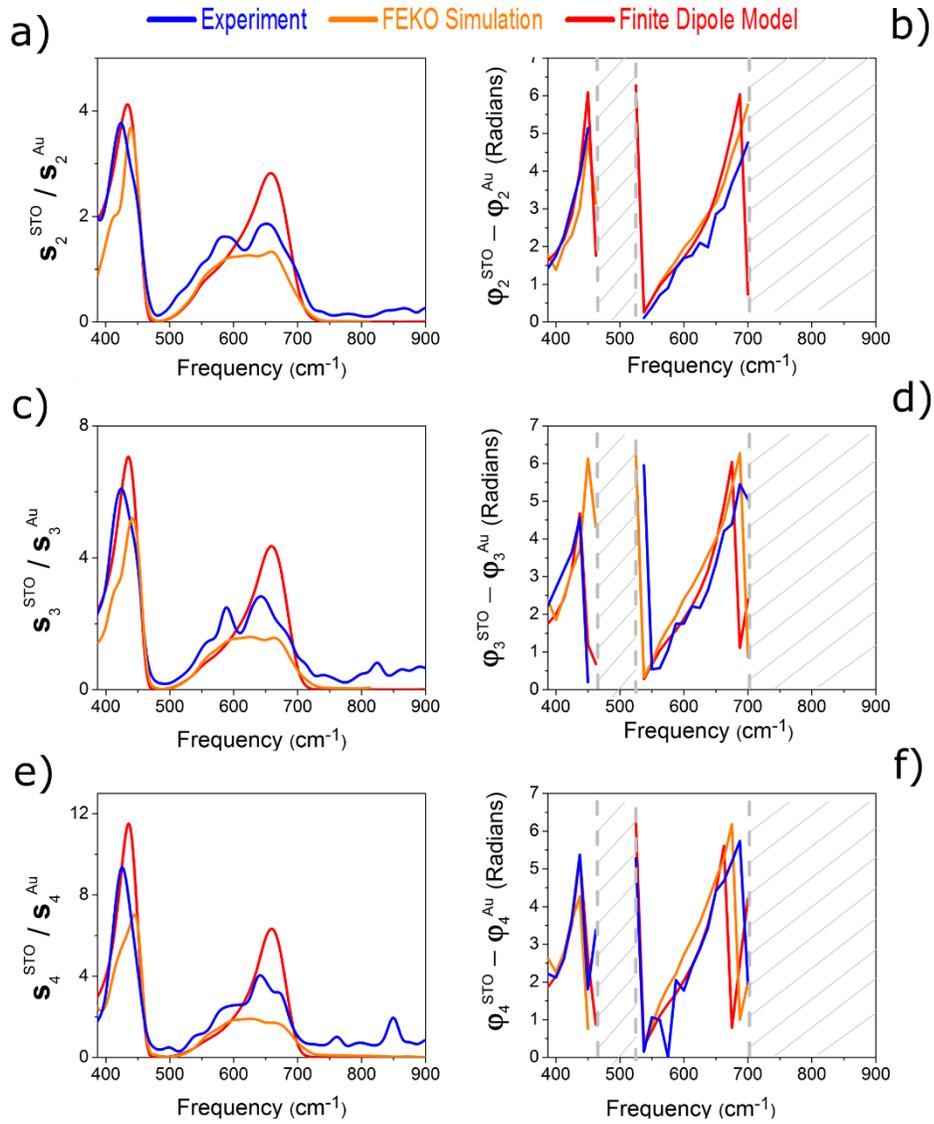

**FIG. 9.** (a-f) Plots show the experimental results, numerical simulations, and finite dipole model calculations of the n = 2, 3, 4 near-field infrared amplitude and phase of STO normalized to the spectra on gold. The parameters (L, $g_1$, $g_2$) for the finite dipole model were obtained from set 1 of Fig. 7.

**Appendix B: Antenna Resonances**

By simulating the detailed probe geometry, we observe that the simulated scattered far- and near-field signals exhibit antenna resonances [48]. These resonances correspond to charge oscillations of either even or odd charge symmetry at the antenna's end [49], and are plasmonic in origin, coming from the metallic layer of our AFM tip. That is, for every number of odd half wavelengths, interference along the shaft of the antenna produces opposite charge accumulation at the antenna's ends, constituting a net electric dipole moment. For an even number of wavelengths, the charge is symmetric at the ends, and so the net dipole moment is zero for these modes. An incident electric field aligned with the tip axis will only excite the odd resonances or bright modes. Nonoblique incidence can excite the even or dark modes through field retardation along the probe axis [56]. To characterize the spectral response of our model we simulated the scattered near- and far-field signals from the entire probe structure without a sample underneath the tip. We computed the electric field distribution 10 nm underneath the tip's apex and compared this to the scattered far-field. These results are displayed in Fig. 10. The resonant structure is evident in both the far- and near-field spectra. Since we are illuminating our probe at an oblique angle relative to the tip axis, both bright and dark modes are active. For the bright dipolar modes, the far-field spectrum is blueshifted from the corresponding near-field spectrum. This behavior is expected for the modes dominated by a dipole moment [57]. The dipole moment of the probe $p(\omega)$ can be modeled as a series of driven damped harmonic oscillators with Lorentzian line shapes. If one inspects the field intensity, $I \propto |E|^2$, in both the far- and near-fields from an electric dipole, a factor of $\omega^4$ appears in the far-field intensity.

$$I_{ff} \propto \omega^4 |p(\omega)|^2 \quad \text{(B1)}$$

$$I_{nf} \propto |p(\omega)|^2 \qquad (B2)$$

For wide resonances in the dipole moment p(ω), the factor of ω⁴ will blueshift the far-field spectrum [57,58]. Dark modes (*l*=2,4 . . .) in the near field have a nontrivial correspondence to the far field because they are not purely dipolar.

It is worth noting the bandwidth of the antenna modes in Fig. 10. In general, plasmonic modes have high losses which result in poor radiation characteristics. The material chosen ultimately provides the upper bound on the quality factor of these resonators [59]. Optimization of the probe geometry allows this upper bound Q factor to be achieved. Work has been done previously, to enhance the Q factors of plasmonic resonators [49,60]. Since our near-field interactions are radiatively coupled to the AFM probe, all of the demodulated spectra will have broadened resonances due to the probe's low Q factor. Future development of high Q factor nanoantennas are an excellent approach to enhance light wavelength confinement, manipulation and transport at the nanoscale.

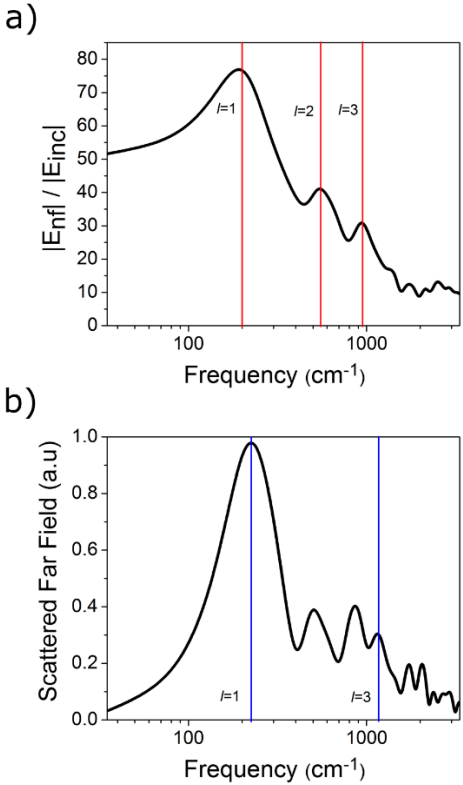

**FIG. 10**. (a) Simulated electric field enhancement 10 nm underneath probe apex without sample and with a p-polarized plane wave incident at angles $\phi = 45°, \theta = 60°$. The vertical lines indicate antenna modes. The *l* = 1 is the fundamental dipolar mode-, *l* = 2 and *l* = 3 are higher order-modes. (b) Far-field spectrum of the probe (without sample) with a p-polarized plane wave incident at angles indicated in (a).

**Appendix C: Angular dependence of scattering**

We simulated the angular distribution of the far-field scattering from the AFM probe. Predominantly forward scattering and backscattering were observed. We also observed a detailed lobed far-field pattern [Fig. 11(a)] which has been seen previously in Ref. [50]. We find that the detailed lobed pattern depends on the incident light frequency. We used the nano-FTIR probe model in the simulations. The radiation patterns of our probe at different frequencies in contact with a bulk Au and STO substrate are shown in Fig. 11. The radiation patterns plotted in Figs. 11(b) and 11(c) are within the plane of incidence of the plane wave which is incident at angles $\phi = 45°$ and $\theta = 60°$. For Au, each radiation pattern corresponds to the peak frequency of the antenna modes identified in Fig. 10(a). As plotted in Fig. 11(b), a dipolar radiation pattern is observed for the $l=1$ antenna mode. As the frequency is increased the lobe structure emerges with an enhancement in forward scattering. The main radiation lobes' angular bandwidth decreases with increasing frequency which also introduces additional side lobes, as expected from antenna theory [61]. In Fig. 11(c) the far-field radiation patterns of the probe in contact with STO near its SPhP resonance frequencies are plotted. Similar lobe structure exists for Au and STO, except for a slight increase in the backscattered lobe's angular bandwidth for STO.

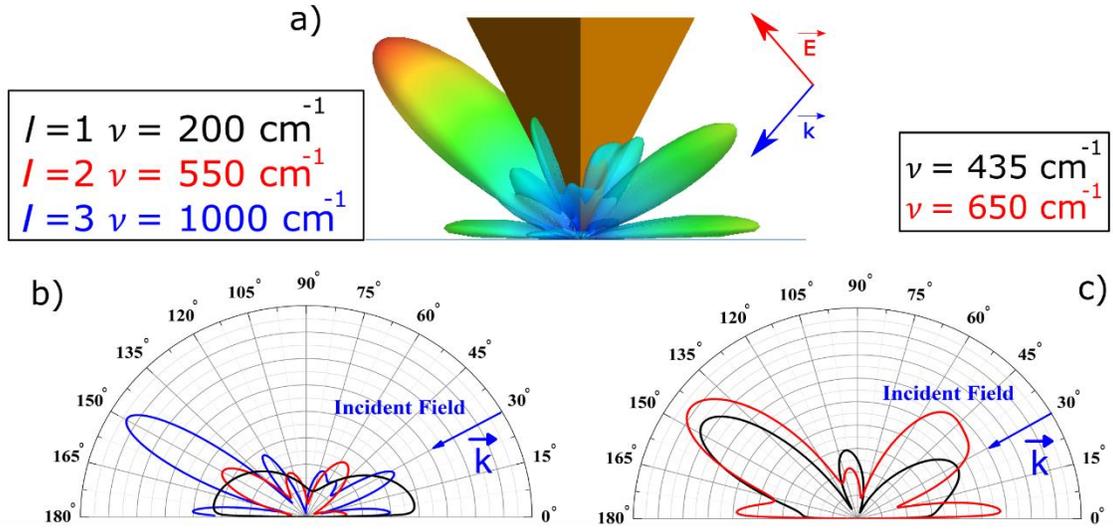

**Figure 11**. (a) Three dimensional far-field scattering of our probe-sample model for the tip in close proximity to gold. (b) Plot of the scattered electric field as a function of the angle $\theta$ along the incidence plane ($\phi = 45°, \theta = 0° - 180°$) at the first three antenna mode frequencies listed above the plot obtained from Fig 10(a). At the fundamental dipolar mode $l=1$, the probe exhibits a dipole pattern. For each higher order antenna mode $l=2,3, \ldots$ the main scattering lobe sharpens, and as a consequence additional side lobes emerge. (c) Plot of the scattered electric field as a function of the angle $\theta$ for the same incidence plane as (b) but for frequencies 435 and 650 cm$^{-1}$ that are near SPhP resonance frequencies of STO.

**Appendix D: Discussion of the dip at ~ 610 cm$^{-1}$ in the experimental STO/Au amplitude spectra**

As stated in the main text, a dip at ~ 610 cm$^{-1}$ in the experimental STO/Au amplitude spectra is present for the n=2,3 harmonics. This dip leads to a double-peak structure in the spectra that is not captured by the numerical results presented in the manuscript. Moreover, this dip is not captured by the point dipole model and the finite dipole model. We hypothesize the dip is either due to partial destructive optical interference or due to the anharmonic phonon effect in STO not captured by the published dielectric function. At present, we consider two possibilities for the partial destructive optical interference: it may arise from a convergent incident beam and/or from the cantilever of the AFM probe. This hypothesis is supported by the results of two distinct sets of numerical simulations. One simulation was performed with a convergent incident beam with a diffraction limited spot at the tip's apex and using the published values of the STO dielectric function. Another simulation was performed with an incident plane-wave using published values of the STO dielectric function and a modified AFM probe model which includes a partial cantilever. At present, the results are not conclusive because the complexity of the above simulations leads to uncertainties that are comparable to the relatively small dip feature we are attempting to capture. The second hypothesis for the origin of the dip feature is supported by our numerical simulations performed with an incident plane wave but with slightly modified values of the STO dielectric function that account for the anharmonic phonon effect near ~ 620 cm$^{-1}$ [62,63]. This hypothesis is also supported by previous work which suggests that phonon anharmonicity leads to a feature at ~ 620 cm$^{-1}$ in the published far-field reflection measurements [62,63].

**Appendix E: Dielectric functions used for simulations**

In Fig. 12, we plot the frequency dependent real ($\varepsilon_1$) and imaginary ($\varepsilon_2$) parts of the dielectric functions of materials used in the simulations [62,64-67].

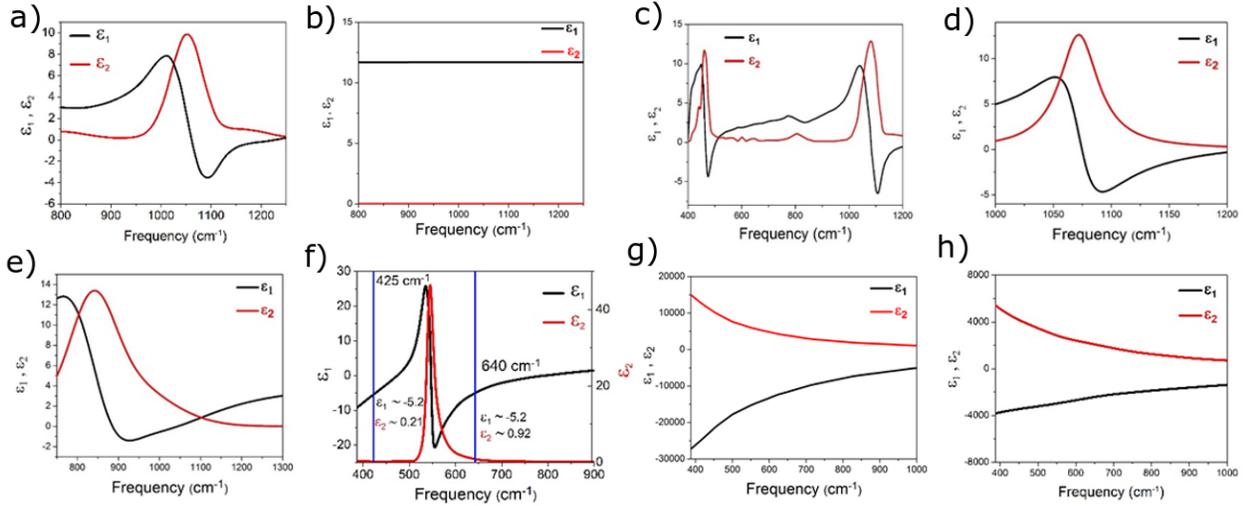

**FIG 12**. (a) The dielectric function of amorphous $SiO_2$ taken from Ref.[64] and used to obtain the simulation results in Fig 3. (b) Dielectric function of Si, taken from Ref. [64] and used to obtain the simulation results in Fig. 3, 4(b)-,4(c), and 6. (c) Dielectric function of amorphous $SiO_2$, taken from Ref. [65] and used to obtain the simulation results in Figs. 4(a) -4(c). (d) Dielectric function of amorphous $SiO_2$, taken from Ref. [45] and used to obtain the simulation results in Fig. 4(d). (e) Dielectric function of amorphous $Si_xN_y$, taken from Ref. [66] and used to obtain the simulation results in Fig. 4(e). (f) Dielectric function of STO taken from Ref. [62] and used to obtain the simulation results in Figs. 5 and 6. Vertical lines indicate frequencies of the observed experimental SPhP resonance peaks. g) Dielectric function of Au, taken from Ref. [67] and used to obtain the simulation results in Figs. 5 and 6.( h) Dielectric function of Pt, taken from Ref. [67] and used to obtain the simulation results in Fig. 6.